\documentclass[12pt]{article}
\usepackage{amsmath,epsfig,graphicx,amssymb}

\newcommand{\beq}{\begin{equation}}
\newcommand{\eeq}{\end{equation}}
\newcommand{\ben}{\begin{eqnarray}}
\newcommand{\een}{\end{eqnarray}}
\newcommand{\bes}{\begin{subequations}}
\newcommand{\ees}{\end{subequations}}
\newcommand{\bFig}{\begin{figure}}
\newcommand{\eFig}{\end{figure}}

\begin{document}

\title{Measurement As Spontaneous Symmetry Breaking, Non-locality and Non-Boolean Holism}
\author{Partha Ghose\footnote{partha.ghose@gmail.com} \\
Centre for Astroparticle Physics and Space Science (CAPPS),\\Bose Institute, \\ Block EN, Sector V, Salt Lake, Kolkata 700 091, India}
\maketitle
\begin{abstract}
It is shown that having degenerate ground states over the domain of the wavefunction of a system is a sufficient condition for a quantum system to act as a measuring apparatus for the system. Measurements are then instances of spontaneous symmetry breaking to one of these ground states, induced by environmental perturbations. Together with non-Boolean holism this constitutes an optimal formulation of quantum mechanics that does not imply non-locality. 

\end{abstract}
\vskip 0.1in
Keywords: measurement paradox, spontaneous symmetry breaking, non-Boolean holism, nonlocality
\section{Introduction}
Measurement has been a longstanding paradox in quantum mechanics when regarded as a theory applicable to all physical systems with no arbitrary split between the observer and the observed. To make a measurement on a quantum system $S$ described by a wavefunction $\psi (S)$, one has to couple it to a measuring apparatus $M$, a macroscopic system, with a wavefunction $\psi (M)$ and allow them to evolve to the (normalized) entangled state
\beq
\Psi (SM) = \sum_{m, n} c_{m, n} \psi_m (S) \psi_n (M) \label{ent}
\eeq
where $\{\psi_m (S)\}$ and $\{\psi_n (M)\}$ are a complete set of orthonormal basis wavefunctions for the system and the measuring apparatus respectively. This corresponds to the pure density matrix
\beq
\rho = \Psi (SM) \Psi^* (SM) \label{pure}
\eeq
To account for the definite results one observes in practice, the measurement axiom requires one to delete all the terms in (\ref{ent}) and retain only {\em one} of them, say $\psi_m (S)\psi_m (M)$, such that $\rho$ is diagonal with a single non-vanishing diagonal element, and a reading of the apparatus state $\psi_m (M)$ leads to a determination of the uniquely correlated system state $\psi_m (S)$. This leads to the controversial process of `collapse' which von Neumann \cite{vN} described as a non-unitary `projection' from the pure state (\ref{pure}) to a mixed state described by $\rho \rightarrow \tilde{\rho} = \sum_m \Pi_m \rho \Pi_m$ with $\Pi_m = (\psi_m (S) \psi_m (M))(\psi_m (S) \psi_m (M))^*$ with $\sum_m\vert c_{m, m}\vert^2 = 1$ and $\vert c_{m, m}\vert^2$ the probabilities of the possible outcomes. The matrix $\tilde{\rho}$ is called the `reduced density matrix'. To interpret the resulting state $\psi_m (S)$ as a wavefunction, one has to normalize it. This non-unitary projection is clearly inconsistent with unitary Schr\"{o}dinger evolution, and this has been the `central mystery' of quantum mechanics \cite{ghose}. As Schr\"{o}dinger \cite{schr} put it,
\begin{quote}
$\cdots$ any {\em measurement} suspends the law that otherwise governs continuous time-dependence of the $\psi$-function and brings about in it a quite different change, not governed by any law but rather dictated by the result of the measurement.
\end{quote}

It is important clearly to understand the nature of the problem and the broad classes of viewpoints that have been expressed to deal with it. There is the widely held viewpoint that there is no inner contradiction in interpreting measurement as a quantum mechanical process. On the other hand, there is the view that `measurement may well be explained by quantum theory in the sense that ``quantum-mechanical noncausality'' can be derived from statistical uncertainties inherent in the necessarily macroscopic apparatus of measurement' or the larger environment \cite{zeh}. Zeh has claimed that neither of these arguments is valid. The first viewpoint is flawed, in his opinion, because there is no dynamical mechanism within quantum theory for the non-occurrence of superpositions of macroscopic states (the Schr\"{o}dinger cat paradox). This is indeed true for isolated or closed systems, but the only such system is the universe itself. The second viewpoint uses a circular argument because it makes use of the density matrix formalism which is itself based on the axiom of measurement. An ensemble cannot be derived from the density matrix. For example, although different statistical ensembles consisting of equal probabilities of silver atoms with spins up and down along different spatial directions can be prepared using the Stern-Gerlach method, they are described by the same density matrix as long as the axiom of measurement is accepted. Hence, the density matrix formalism cannot be a complete description of a statistical ensemble. Although decoherence theory \cite{zurek} belongs to this class, it has its own advantages and importance in this context which will become clear in what follows. Other viewpoints also exist, mostly introducing additional physical concepts, and references to them can be found in Ref. \cite{ghose}. 

\section{Measurement As Spontaneous Symmetry Breaking}

A satisfactory resolution of the measurement problem requires that it be viewed as a quantum mechanical process in conformity with the first viewpoint mentioned above. It turns out that it {\em can} be so viewed, namely as a unitary and spontaneous transition from one quantum state to another, somewhat analogous to (but not identical with) spontaneous decays of atomic and other states. Let us, for example, consider the typical Stern-Gerlach measurement of a spin-$1/2$ atom $S$. Let 
\beq
\vert X\rangle = [a \vert \psi_S\rangle\vert\uparrow\rangle_z\vert \psi_M\rangle\vert_{p(1)} + b \vert \psi_S\rangle\vert\downarrow\rangle_z\vert \psi_M\rangle\vert_{p(2)} ]e^{i(kx - \omega t)}\label{X} 
\eeq
be the entangled state of the atom-detector system at time $t$ with $\vert a\vert^2 + \vert b\vert^2 = 1$. Here $p(1)$ and $p(2)$ denote the macroscopically distinct supports of the spatial wavefunctions $\psi_S$ and $\psi_M$ of the spin and detector systems in the $z > 0$ and $z < 0$ regions of the $xz$ plane respectively. When treated as a closed system, this entangled state is stable acording to quantum mechanics. But the discrete events that are known to occur at the detectors suggest that it becomes unstable when subjected to unavoidable perurbations due either to its immediate environment or intrinsic quantum fluctuations, and it makes a spontaneous transition to either the state
\beq
\vert \psi_S\rangle\vert\uparrow\rangle_z\vert \psi_M\rangle\vert_{p(1)}e^{i(kx - \omega t)}
\eeq
with probability $\vert a\vert^2$ or the state
\beq 
\vert \psi_S\rangle\vert\downarrow\rangle_z\vert \psi_M\rangle\vert_{p(2)}e^{i(kx - \omega t)}  
\eeq
with probability $\vert b\vert^2$. In general, observations suggest that a closed system 
\beq
\vert X\rangle = \sum_{m, n} c_{m, n}\vert S_m\rangle \vert M_n \rangle \label{x}
\eeq
in the Hilbert bundle describing the entangled system-measuring apparatus state becomes unstable against small perturbations and spontaneously orients itself along one of the possible base rays $\vert M_m\rangle \vert S_m\rangle$ with probability $\vert c_{m, m} \vert^2$. {\em Since the ray orients itself along one of the base rays, its projections on all the other base rays vanish}. The principal difference from the conventional projection postulate is therefore simply this: whereas in the conventional case, a measurement result corresponds to a non-unitary `projection' of the total ray to one of its possible component base rays which then has to be re-normalized again (a patently non-quantum mechanical process), according to the new viewpoint the ray spontaneously orients itself completely along this ray with the same probability (a purely quantum mechanical process). Unitarity is preserved in the process, and the components along all other base rays vanish, unlike in the conventional case.

These transitions can be viewed as instances of spontaneous symmetry breaking (SSB) in the following sense. Let all states of the measuring apparatus (the ``pointer states'') before measurement be {\em a priori} equally probable. Then they can be viewed as degenerate ground states \cite{zimanyi}. The different possibilities given by the right-hand side of Eqn. (\ref{x}) can be viewed as ``attractors'' of the entangled state. When arbitrarily small and uniform perturbations are present, the state can make a unitary transition (by Schr\"{o}dinger evolution) to one of these attractors, the choice of the particular attractor in any given event being by pure chance. Once the state reaches one of these degenerate attractors, it becomes stable because there is no dynamical reason for it to shift to any other attractor. The different probabilities of transition to these attractors are determined by the entangled state (\ref{x}) which breaks the symmetry of the pointer states.  For example, in the case of the Stern-Gerlach set up, before measurement the two detectors are designed to be degenerate in the sense that {\em a priori} they are equally likely to click, and the entangled state determines the probabilities with which they click at random. 
Another example is the double-slit interference pattern produced by a particle beam. Every scintillation on the final screen breaks the designed homogeneity (i.e., translation symmetry) of the screen before the measurement, and the entangled state determines the final statistical pattern.  

It is clear therefore that {\em any quantum system $M$ which has degenerate ground states (i.e., states of equal a priori probability) over the domain of the wavefunction of $S$ before measurement can act as a measuring apparatus for $S$}. It is, of course, much easier in practice to use macroscopic systems to design such apparatuses of arbitraily high precision than to use microscopic systems, but in principle microscopic systems with the required degeneracy property can also be used. 

What remains to be shown is that entangled systems become unstable against small external perturbations. Decoherence theory \cite{zurek} provides this required theoretical support. Decoherence theory shows that when a pure quantum system like $X$ interacts and gets entangled with its environment (modelled as a heat bath of quantum oscillators), its density matrix becomes rapidly diagonal in the pointer state basis. What decoherence does not explain is explained by spontaneous symmetry breaking of the type advocated above, namely the occurrence of a single term of the diagonal matrix in individual events. 

Some fundamental differences from usual cases of SSB in statistical mechanics and high energy physics must be emphasized. They are all examples of deterministic SSB. The symmetries in all these cases are broken `at the level of probabilistic distributions rather than at the level of chancy events' \cite{liu}. On the other hand, the transition to a particular attractor state in quantum measurement is a purely chance event. The probability that a particular attractor state occurs is not determined by which pointer state is the `most probable' (they are all equally probable by choice) but by the structure of the entangled state before measurement.

\section{Quantum Non-locality?}

It is principally due to Bell's pioneering work \cite{bell} that there is now more or less a concensus of opinion that quantum mechanics is incompatible with `local realism', i.e. the requirements of locality + realism. A {\em local realist theory} is usually understood to satisfy the following criteria:

\begin{enumerate}

\item {\em Realism}:  Objects have definite values of all measurable properties, such as position, momentum and spin, prior to and independent of measurement.

\item {\em Locality}: Consider two spatially separated objects $A$ and $B$ which are non-interacting. Then,  local actions on $A$ can only influence $A$ and not $B$. 
\end{enumerate}
It is important to note that $A$ and $B$ are, by hypothesis, {\em spatially} separated and {\em non-interacting}, which means there is no potential between them, and hence a local action on $A$ can {\em never ever} influence $B$. Any mutual influence between them predicted by a theory, whether instantaneous or not, must then be considered `spooky' or `telepathic'. The locality condition is, however, often stated in terms of micro-causality, i.e. a local action on $A$ can influence $B$ only through a physical signal that cannot travel faster than light. But this presumes that $A$ and $B$ {\em can} interact and influence each other in the normal way, missing the whole point. In checking non-locality experimentally, of course, it is imperative to eliminate all possible spurious signals from $A$ to $B$.
 
Bell showed that no physical theory of Local Hidden variables (LHV) which satisfies these two criteria can reproduce all of the predictions of quantum mechanics. Many experiments have been done since Bell's paper which claim to rule out LHV theories \cite{ghose2}, the first most convincing one being that of Aspect and his collaborators \cite{aspect}. 

All this was, of course, stimulated by the famous 1935 paper of Einstein, Podolsky and Rosen \cite{EPR} which claimed to establish the {\em incompleteness} of quantum mechanics. That the EPR analysis of quantm mechanics also implied quantum {\em entanglement} was first pointed out by Schr\"{o}dinger \cite{schr}, as we have seen.
Although much has been written and debated on these papers, particularly the definition of `elements of reality' that appears in the EPR paper, it has not been sufficiently emphasized that Einstein himself was not very happy with this version of his views. He wrote to Schr\"{o}dinger on June 19, 1935 \cite{fine}:

\begin{quote}
For reasons of language this [paper] was written by Podolsky after several discussions. Still, it did not come out as well as I had originally wanted; rather, the essential thing was, so to speak, smothered by the formalism [Gelehrsamkeit].
\end{quote}
Let us see how Einstein himself described the `essential thing' much later in his Autobiography \cite{einstein-auto} which can be taken as his considered and final view of the matter:
\begin{quote}
For the further discussion I shall assume two physicists, $A$ and $B$, who represent a different conception with reference to the real situation as described by the $\Psi$-function.

$A$. The individual system (before the measurement) has a definite value of $q$ (i.e. $p$) for all variables of the system, and more specifically, {\em that} value which is determined by a measurement of this variable. Proceeding from this conception, he will state: The $\Psi$-function is no exhaustive description of the real situation of the system but an incomplete description; it expresses only what we know on the basis of former measurements concerning the system.

$B$. The individual system (before measurement) has no definite value of $q$ (i.e., $p$). The value of the measurement only arises in cooperation with the unique probability which is given to it in view of the $\Psi$-function only through the act of measurement itself. Proceeding from this conception, he will (or, at least, he may) state: the $\Psi$-function is an exhaustive description of the real situation of the system.

We now present to these two physicists the following instance: There is to be a system which at the time $t$ of our observation consists of two partial systems $S_1$ and $S_2$, which at this time are spatially separated and (in the sense of the classical physics) are without significant reciprocity. The total system is to be completely described through a known $\Psi$-function $\Psi_{12}$ in the sense of quantum mechanics. All quantum theoreticians now agree upon the following: If I make a complete measurement of $S_1$, I get from the results of the measurement and from $\Psi_{12}$ an entirely definite $\Psi$-function $\Psi_2$ of the system $S_2$. The character of $\Psi_2$ then depends upon {\em what kind} of measurement I undertake on $S_1$.

Now it appears to me that one may speak of the real factual situation of the partial system $S_2$. Of this real factual situation, we know to begin with, before the measurement of $S_1$, even less than we know of a system described by the $\Psi$-function. But on one supposition we should, in my opinion, absolutely hold fast: the real factual situation of the system $S_2$ is independent of what is done with the system $S_1$, which is spatially separated from the former. According to the type of measurement which I make of $S_1$, I get, however, a very different $\Psi_2$ for the second partial system $(\varphi_2, \varphi_2^1, \cdots)$. Now, however, the real factual situation of $S_2$ must be independent of what happens to $S_1$. For the same real situation of $S_2$ it is possible therefore to find, according to one's choice, different types of $\Psi$-function. (One can escape from this conclusion only by either assuming that the measurement of $S_1$ ((telepathically)) changes the real factual situation of $S_2$ or by denying independent real situations as such to things which are spatially separated from each other. Both alternatives appear to me entirely unacceptable.)

If now the physicists, $A$ and $B$, accept this consideration as valid, then $B$ will have to give up his position that the $\Psi$-function constitutes a complete description of a real factual situation. For in this case it would be impossible that the two different types of $\Psi$-functions could be co-ordinated with the identical factual situation of $S_2$.
\end{quote}

To critically assess what Einstein is saying it is important to bear in mind two theorems concerning wavefunctions that Schr\"{o}dinger enunciated \cite{schr}:
\begin{quote}
Theorem 1: If different $\psi$-functions are under discussion the system is in different states.

Theorem 2: For the same $\psi$-function the system is in the same state.
\end{quote}

With this background Einstein's statement may be critiqued as follows. Let us consider measurements of two non-commuting observables (like position and momentum) on $S_1$. According to quantum mechanics, one must write the wavefunction of the total system in two different bases in the forms
\ben
\Psi_{12} &=& \sum_{m, n = 1}^N c_{m, n} \psi_m (S_1) \Psi_n (S_2)\label{1}\\
&=& \sum_{m, n = 1}^N d_{m, n} \phi_m (S_1) \varphi_n (S_2)\label{2} 
\een
where $\{\psi_m (S_1), \Psi_n (S_2)\}$ and $\{\phi_m (S_1), \varphi_n (S_2)\}$ are complete sets of orthonormal bases for $S_1$ and $S_2$ appropriate to the two non-commuting observables to be measured. Hence, different kinds of measurement on $S_1$ correspond to different wavefunctions $\Psi_n (S_2)$ and $\varphi_n (S_2)$ of $S_2$. But, according to Einstein, different wavefunctions for $S_2$ depending on what one {\em chooses} to measure on $S_1$ is inconsistent with the same `real factual situation' of $S_2$. Hence, the quantum mechanical description of the real factual situation of $S_2$ is incomplete.

The standard reaction of quantum physicists to this would be that the `real factual situation' of a system and its independence of what is done to spatially separated systems is precisely equivalent to `local realism' of the Bell type which is now known to be incompatible with quantum mechanics. 

\section{Non-Boolean Holism}

This view is actually based on a false imposition of reductionism or Boolean logic on quantum mechanics which describes physical reality in terms of a {\em transitive partial Boolean algebra} \cite{primas}.
Primas has emphasized that it is advantageous to start with a holistic or undivided universe in which there are no {\em a priori} given part-whole distinctions. In such a universe the existence of holistic correlations between contextually chosen parts is intrinsic and natural. Primas calls this type of holism {\em non-Boolean holism} as opposed to {\em Boolean holism} in which the whole consists of parts though it is more than the sum of the parts.

In an undivided entity there are no {\em a priori} given part-whole distinctions. It is only when a division is made that correlations between the created parts emerge as natural consequences, much like the coastline correlations that emerged when the continents broke away from Pangea due to plate tectonics. The very word `correlation' presupposes the existence of parts -- in an undivided entity without parts there are no correlations. The same undivided entity can, of course, be partitioned or divided in many ways, giving rise to different correlations in every case. 

Now, the concept of entanglement was first introduced by Schr\"{o}dinger \cite{schr} in his famous 1935 paper in which he wrote:
\begin{quote}
If two separated bodies, each by itself known maximally, enter a situation in which they influence each other, and separate again, then there occurs regularly that which I have just called {\em entanglement} of our knowledge of the two bodies. $\cdots$

Best possible knowledge of a whole does {\em not} include best possible knowledge of its parts -- and that is what keeps coming back to haunt us.
\end{quote}
The haunting arises from Boolean holism. An entangled state is, however, non-Boolean in the sense that the partial systems lose their identity completely, and this must be accepted if the theory is to make sense.
Hence, the observed quantum correlations between the partial systems $M$ and $S$ (or between $S_1$ and $S_2$), which have no existence of their own in the entangled state, must be {\em locally created} in the process of measurement which is an act of `conditional disjunction' -- `if the mark is at $1$, {\em then} things are thus and so for the measured object, {\em if} it is at line $2$, then such and such, if at $3$, then a third, etc.' \cite{schr}. Schr\"{o}dinger calls this a {\em conditional disjunction} of the total or entangled wavefunction. Hence, {\em quantum correlations are
not produced by `telepathic' effects between already existing and independent distant objects}.

Thus, instead of being a nagging problem, measurement can be recognized as the very process of creation of parts from the whole, or the many from the one. 
  
Einstein was right in concluding from his analysis that the only two options available were telepathy and non-separability, but unfortunately he rejected both, though, as we have seen, non-separability is integral to quantum mechanics. On the other hand, many of today's physicists seem to have no qualms in accepting telepathy. 

\section{Conclusions}

The main interpretative problems disappear once it is realized that (a) the real objective world reflected in the Hilbert bundle is globally non-Boolean but locally Boolean and that (b) measurements are instances of spontaneous symmetry breaking in the sense defined above. As Primas points out:
\begin{quote}
The nonseparability predicted by quantum theory is usually described
by entanglement, a term introduced by Erwin Schr¨odinger \cite{schr}. His historical
characterization still adopts an atomistic ontology, assuming that
the quantum world consists of parts. From the modern viewpoint it is
therefore somewhat misleading to speak of ``an entanglement of quantum
systems'', since subsystems have no independent existence. Moreover,
genuine holistic correlations are not restricted to physical systems. They
are independent of Planck's constant of action, they are independent of
spatial separations, and do not arise from known physical forces. They
cannot be reproduced by any system with a Boolean logical structure. In
systems which allow a Boolean description there are no entanglements.
\end{quote}

\section{Acknowledgement}
The essential ideas contained in this paper were presented at the International Conference on ``Quantum Systems, Classical Measurements and Consequences'' at SVYASA University, Bangalore on June 28, 2010 and at the Indian Institute of Science, Bangalore on July 1, 2010. I would like to thank A. Patel and N. D. Haridass for helpful discussions after the talk at IISc and for pointing out Ref. \cite{zimanyi} to me.
I thank the National Academy of Sciences, India for the award of a Senior Scientist Platinum Jubilee Fellowship which allowed this work to be undertaken.


\begin{thebibliography}{0}
\bibitem{vN}
J. von Neumann, Mathematical Foundations of Quantum Mechanics, Princeton University Press, 1955. 
\bibitem{ghose}
P. Ghose, {\em Neuroquantology} {\bf 7} (2009)623-634, ; ArXiv: 0906.0898 v1 [quant-ph] 4 June 2009.
\bibitem{schr}
E. Schr\"{o}dinger, {\em Proc. of the American Phys. Soc. } {\bf 124} (1935) 323-38. Reprinted in {\em Quantum Theory of Measurement}, eds. J. A. Wheeler \& W. H. Zurek, Princeton University Press, New Jersey, 1983.
\bibitem{zeh}
H. D. Zeh, {\em Found. of Phys.} {\bf 1} (1970) 69-76 and references therein.
\bibitem{zurek}
W. H. Zurek (2002).
http://arxiv.org/pdf/quant-ph/0306072v1 and references therein. 
\bibitem{zimanyi}
G. T. Zimanyi and K. Vladar, {\em Found. of Phys. Lett.} {\bf 1} (1988) 175-185.
\bibitem{liu}
Chuang Liu, {\em Phil. of Sc.} {\bf 70} (2003) 590-608; philsci-archive@mail.pitt.edu, footnote 2.
\bibitem{bell}
J. S. Bell, {\em Physics} {\bf 1} (1964) 195-200. 
\bibitem{ghose2}
P. Ghose (1999). Testing Quantum Mechanics on New Ground, Cambridge University Press, Cambridge, 1964, Chapter 9.  
\bibitem{aspect}
A. Aspect, J. Dalibard and G. Roger, {\em Phys. Rev. Lett.} {\bf 49} (1982) 1804-1807.
\bibitem{EPR}
A. Einstein, B. Podolsky and N. Rosen, {\em Phys. Rev.} {\bf 37} (1935) 780-81.
\bibitem{fine}
A. Fine, The Shaky Game: Einstein, Realism and the Quantum Theory, 2nd Edition, University of Chicago Press, Chicago, 1996, p. 35.
\bibitem{einstein-auto}
A. Einstein, Autobiographical Notes in {\it Albert Einstein: Philosopher-Scientist}, edited by P. A. Schilpp (The Library of Living Philosophers, Evanston), 1949, pp. 83-87.
\bibitem{primas}
Primas H, {\em Mind \& Matter} {\bf 5}(1) (2007) 7-44. 

\end{thebibliography}
\end{document}